\documentclass[11pt]{article}
\pdfoutput=1
\usepackage{amsfonts,amssymb,amsmath,epsfig, graphicx,yfonts,jheppub}%
\usepackage[utf8]{inputenc}
\usepackage{lineno}
\usepackage{cancel}
\usepackage{subcaption}
\usepackage[normalem]{ulem}

\def\sc{\ensuremath{^{\mathsf{c}}}}

\subheader{\begin{flushright}
\texttt{}
\end{flushright}}

\title{Position dependence of the holographic entanglement entropy for an accelerating quark-antiquark  pair}

\author{Andrés Argandoña}
\author{and Alberto G\"uijosa}
\affiliation{Departamento de Física de Altas Energías, Instituto de Ciencias Nucleares, \\ 
Universidad Nacional Aut\'onoma de M\'exico, Apdo. Postal 70-543, CdMx 04510, M\'exico}

\abstract{Through the holographic correspondence, we compute the entanglement entropy of the gluonic field sourced by a quark-antiquark pair undergoing uniform back-to-back acceleration. Previous calculations had obtained this only for the case where the entanglement surface is located midway between the quark and antiquark. Here, we consider the more general case with a relative lateral displacement, and determine the entanglement entropy as a function of the distance between the quark and the entanglement surface.
This setup is of interest because it departs from the usual simplifying conditions of staticity and thermality, and because it yields more information about the entanglement pattern in the gluonic field and about the possibility of eventually developing a purely worldsheet interpretation for said entanglement. 
}

\emailAdd{andres.argandona@correo.nucleares.unam.mx, alberto@nucleares.unam.mx}

\begin{document} 

\maketitle
\setlength{\parskip}{3pt}

\section{Introduction and Summary}\label{introduction}
When the study of the holographic correspondence \cite{Maldacena:1997re,Gubser:1998bc,Witten:1998qj} began, part of its importance lay in the fact that it allowed the characterization of interesting aspects of strongly coupled gauge theories. A standard way to analyze such theories is by studying their response to external charges, whose trajectories define Wilson lines in all possible representations \cite{Maldacena:1998im,Rey:1998ik,Gomis:2006sb}. The profile of the gluonic and other fields sourced by these charges can be mapped out by correlators with local operators \cite{Danielsson:1998wt,Callan:1999ki}. This set of tools has yielded many valuable lessons, including those reviewed in \cite{Sonnenschein:1999if,Semenoff:2002kk,Kristjansen:2010kg,Alday:2010kn,Chernicoff:2011xv,Zarembo:2016bbk}.

In a parallel development, over the past couple of decades it has proven very fruitful to
study field theories by inquiring into the quantum entanglement they give rise to \cite{Bombelli:1986rw,Srednicki:1993im,Callan:1994py,Holzhey:1994we,Osterloh:2002sym,Osborne:2002zz,Vidal:2002rm,Peschel_2003,Calabrese:2004eu,Casini:2004bw,Barthel:2006ct,Casini:2009sr,Casini:2022rlv}. In particular, computations of entanglement entropy (EE) associated with spatial regions give  direct access to the pattern of entanglement that underlies nonvanishing correlators of local operators \cite{Wolf:2007tdq}. Such computations thus provide an alternative route to examine the effect of external charges in gauge theories. The holographic prescription for calculating EE was conjectured originally by Ryu and Takayanagi (RT) in 
\cite{Ryu:2006bv,Ryu:2006ef} and proved later in \cite{Casini:2011kv,Lewkowycz:2013nqa}. Its range of applicability has been progressively expanded in various directions \cite{Hubeny:2007xt,Faulkner:2013ana,Dong:2013qoa,Camps:2013zua,Engelhardt:2014gca,Dong:2016hjy}, leading to many interesting results \cite{VanRaamsdonk:2016exw,Rangamani:2016dms,Nishioka:2018khk,Headrick:2019eth}.
Pioneering applications to determine EE in the presence of external charges were carried out in 
\cite{Chang:2013mca,Jensen:2013ora,Jensen:2013lxa,Lewkowycz:2013laa,Karch:2014ufa}.

A configuration that has garnered significant attention \cite{Xiao:2008nr,Caceres:2010rm,Chernicoff:2010yv,Garcia:2012gw,Jensen:2013ora,Sonner:2013mba,Chernicoff:2013iga,Lewkowycz:2013laa,Jensen:2014lua,Hubeny:2014kma,Hubeny:2014zna,Chen:2016xqz,Murata:2017rbp,Kawamoto:2022isn,deBoer:2022rbn,Grieninger:2023ehb,Yeh:2023avs,Grieninger:2023pyb}, and that will also be the focus of the present paper, is a quark and antiquark that undergo back-to-back uniform acceleration  in the vacuum of a $(3+1)$-dimensional conformal field theory (CFT), such as maximally supersymmetric Yang-Mills (MSYM).  
In the dual bulk description, this corresponds to a U-shaped string in a pure anti-de Sitter (AdS) background, with endpoints located on the AdS boundary and moving also with constant acceleration.\footnote{We will restrict attention here to the case where the string extends all the way to the conformal boundary, which corresponds to the quark and antiquark having infinite mass. Interesting novelties arise when the string ends instead on a flavor D-brane at some finite radial depth, which corresponds to the quark and antiquark having a finite mass, and therefore acquiring a finite size \cite{Hovdebo:2005hm,Chernicoff:2009re,Chernicoff:2009ff,Chernicoff:2010yv,Agon:2014rda}.} The string embedding was first obtained in \cite{Xiao:2008nr}, and was later understood \cite{Garcia:2012gw} to be a particular case of the embeddings constructed for arbitrary quark motion in \cite{Mikhailov:2003er}. We will give a more detailed description in the following section, but for now it is important to emphasize that the induced metric on this accelerating string endows the worldsheet with the same causal structure as an eternal AdS$_2$ black hole \cite{Xiao:2008nr}. This 
means that there is a double-sided horizon at a fixed radial depth in AdS, and two distinct  exterior regions, each corresponding to one string endpoint. These exterior regions are connected by a non-traversable wormhole, i.e., an Einstein-Rosen (ER) bridge. This geometric connection can be recognized to arise from the entanglement of the $q$-$\bar{q}$ color-singlet pair, which is an Einstein-Rosen-Podolski (EPR) pair. Altogether, then, the setup gives a concrete example
\cite{Jensen:2013ora} of the ER=EPR conjecture \cite{Maldacena:2013xja}.\footnote{Consideration of more general $q$-$\bar{q}$ trajectories leads to the conclusion that a wormhole emerges if and only if the particles emit gluonic radiation, thereby carrying color away to infinity \cite{Chernicoff:2013iga}.}

\begin{figure}
  \centering
  \includegraphics[width=1\textwidth]{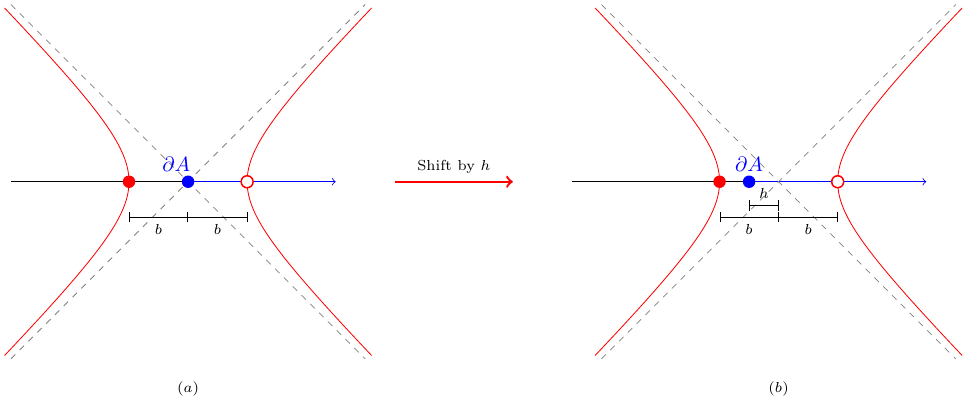}
  \caption{Setup for the computation of entanglement entropy in the presence of a quark (filled red circle) and antiquark (unfilled red circle) that accelerate back-to-back with uniform acceleration $b^{-1}$ and closest distance of approach $2b$.\protect\footnotemark~The red curves are the $q$-$\bar{q}$ worldlines, and the horizontal line is the time slice where the entanglement entropy of the gluonic and other fields is examined. The blue dot indicates the location of the entangling surface $\partial A$, which is a 2-dimensional plane coming out of the picture, with $A$ the spatial region depicted by the blue semi-infinite segment  to the right of $\partial A$. (a) The symmetric configuration, analyzed previously in the literature, where each of the particles is at the same distance from $\partial A$. (b) The asymmetric configuration of interest in the present paper, where $\partial A$ is displaced a distance $h\le b$ to the left of the midpoint between the quark and antiquark.}
  \label{qq shift} 
  \end{figure}

The typical setup involves an entangling surface (ES) that is a plane, splitting a constant-time slice in two, with the uniformly accelerated quark and antiquark positioned symmetrically on either side, equidistant from the plane, as seen in Fig.~\ref{qq shift}a. 
In the present paper, we will be interested in going beyond this highly symmetric scenario, allowing the ES to be positioned at a variable distance from the accelerating quark-antiquark pair, characterized by a parameter $h$, as depicted in Fig.~\ref{qq shift}b.   The motivation for considering this more general setup arises from four  related reasons. First and most directly, it allows us to better map out the pattern of entanglement in the gluonic field sourced by the quark and antiquark. Second, it is to our knowledge the first computation of entanglement associated with a source/defect that \emph{cannot} be reinterpreted as being static in some conformal frame. Third, it is likewise the first computation of source/defect entanglement that does \emph{not} allow a thermal reinterpretation. Fourth and final, having now a movable point of intersection in the bulk between the RT surface and the string worldsheet allows us to ponder whether some meaning can be ascribed to the entanglement calculation directly within the worldsheet perspective.

Let us now discuss each of these motivations in more detail. The first one connects straightforwardly with the opening paragraphs of this Introduction. Just like we would ordinarily probe any given state in a field theory by computing correlators with insertions at arbitrary spatial locations, we would like to compute in said state entanglement entropies with arbitrary entanglement surfaces. As a step towards that goal, we will show in this paper that it is feasible to carry out this computation analytically in the case where the ES is a movable plane.

\footnotetext{A priori this system could have at least three independent parameters: the quark and antiquark accelerations $a_q$ and $a_{\bar{q}}$, together with the $q$-$\bar{q}$ distance at closest approach, $2b$. To select the family of cases where the quark and antiquark Rindler wedges touch at a point, we impose the relation $a_q^{-1}+a_{\bar{q}}^{-1}=2b$. Further restricting to $a_q=a_{\bar{q}}=b^{-1}$, we obtain the only case where on the gravity side the string worldsheet is smooth. For other choices of parameters, additional string segments are involved, which are dual to gluonic shock waves in the gauge theory \cite{Garcia:2012gw}.}

The second and third motivations stem from the fact that the previously studied symmetric configuration (i.e., the case with no displacement, $h=0$) allows for key simplifications. Consider first the CFT in its vacuum state. Starting with Cartesian coordinates where the ES bipartitions the $t=0$ slice at $x^1=0$, one can perform a conformal transformation that maps the reduced density matrix of the causal development of region $x^1 > 0$ to a thermal density matrix on $\mathbb{R} \times \mathbb{H}^{d-1}$ \cite{Casini:2011kv}. This is possible because in this case the modular Hamiltonian is local: it is just the generator of Lorentz boosts in the $x^1$ direction \cite{Bisognano:1975ih,Bisognano:1976za}.  After the conformal map, this operator generates time translations on $\mathbb{R} \times \mathbb{H}^{d-1}$, so the reduced density matrix acquires a thermal form. The same connection continues to hold after we introduce the uniformly accelerated quark and antiquark, as long as they are equidistant from the ES as  in Fig.~\ref{qq shift}a: their hyperbolic worldlines then correspond to orbits of the boost generator, implying that they map to static defects on $\mathbb{R} \times \mathbb{H}^{d-1}$, which in turn ensures that the thermal interpretation applies to the defect system. After Wick rotation, the EE contribution above the vacuum reduces to a thermal computation  determined by the circular Wilson loop \cite{Pestun:2007rz,Erickson:2000af,Drukker:2000rr} and the one-point function of the stress tensor in its presence \cite{Gomis:2008qa,Okuyama:2006jc,Friess:2006fk,Fiol:2012sg, Correa:2012at}. For certain theories, such as MSYM, the exact result is known \cite{Lewkowycz:2013laa}, and at strong coupling it agrees with the holographic computation \cite{Jensen:2013lxa,Jensen:2013ora,Gentle:2014lva},
\begin{equation}
    S
    \equiv S_{q\bar{q}}-S_{\text{vacuum}}
    = \frac{\sqrt{\lambda}}{3}~,\label{undisplacedentropy}
\end{equation}
where $\lambda$ is the 't~Hooft coupling. This expression is finite because the UV divergences cancel out in the subtraction, and it is independent of the particles' acceleration due to conformal invariance.

By means of a second conformal transformation, one can map  the ES to a sphere and the quark-antiquark trajectory to a static worldline at the center of that sphere. In this presentation, the system can be regarded as a defect CFT (DCFT) associated with a one-dimensional defect that preserves an  $SO(2,1)\times SO(d-1)$ subgroup of the original conformal group \cite{Cardy:1984bb,McAvity:1993ue,McAvity:1995zd}. In this context, the extra contribution (\ref{undisplacedentropy}) to the EE coming from the quark and antiquark is a particular example of defect entanglement entropy  \cite{Estes:2014hka,Kobayashi:2018lil}
 $ S_{\text{defect}} \equiv S_{\text{DCFT}}-S_{\text{CFT}}$, where $S_{\text{CFT}}$ denotes the EE of the ambient CFT, and $S_{\text{DCFT}}$ the corresponding quantity in the presence of the defect. In $d=2$ the defect contribution to the entropy is non-increasing along renormalization group (RG) flows between DCFTs, and hence is often taken as a characterization of the degrees of freedom localized at the defect\footnote{In a DCFT in $d=2$ the existence of a monotonic quantity that decreases along the defect RG flow is the content of the $g$–theorem \cite{Affleck:1991tk,Friedan:2003yc,Casini:2016fgb}. Several proposals generalize this idea to higher dimensional spacetimes \cite{Cuomo:2021rkm,Casini:2022bsu} and to defects of higher co-dimension \cite{Nozaki:2012qd,Gaiotto:2014gha,Estes:2014hka,Jensen:2015swa,Kobayashi:2018lil,Wang:2021mdq}.} \cite{Affleck:1991tk,Calabrese:2004eu,Azeyanagi:2007qj,Friedan:2003yc,Sorensen:2009zqb}.

The simplifying features mentioned in the previous two paragraphs are lost when $h\neq 0$, i.e., when one shifts the spatial location of the accelerating quark and antiquark. The meeting point of their respective Rindler wedges is then displaced from the ES, the antiquark Rindler wedge no longer coincides with the causal development of region $x^1>0$, the particles are no longer static on $\mathbb{R} \times \mathbb{H}^{d-1}$, and the reduced state can no longer be mapped to a thermal density matrix. The procedures employed in \cite{Lewkowycz:2013laa} for the gauge theory calculation and in \cite{Jensen:2013ora,Jensen:2013lxa} for the bulk calculation are therefore no longer applicable.

Another notable feature of the symmetric result (\ref{undisplacedentropy}) is that its bulk instantiation originates entirely from the intersection between the string worldsheet and the unbackreacted RT surface \cite{Jensen:2013lxa,Karch:2014ufa}. This intersection turns out to be located at the horizon of the AdS$_2$ black hole, and the defect entropy can be naturally identified with its entropy \cite{Jensen:2013ora,Chernicoff:2013iga,Lewkowycz:2013nqa}. The fourth and final motivation for our work is exploring to what extent the excess  entropy $S$ due to the quark and antiquark admits an intrinsic worldsheet interpretation in more general scenarios. This perspective becomes particularly relevant when one notices that the worldsheet black hole emits Hawking radiation (in the form of fluctuations of the string embedding) once quantum corrections are taken into account \cite{deBoer:2008gu,Son:2009vu,Caceres:2010rm,Chernicoff:2010yv}, hinting at the possibility that the string worldsheet may serve as a lower-dimensional toy model for black-hole evaporation and information loss, much like JT gravity in island constructions\footnote{Several challenges would need to be addressed to make this precise. Most important would be to determine the induced gravity theory that lives on the worldsheet. Moreover, one would need to implement a mechanism by which the Hawking radiation can escape the worldsheet, allowing for the evaporation process to occur.\label{islandfoot}} \cite{Penington:2019npb,Almheiri:2019psf,Almheiri:2019qdq}. In particular, for $d=3$, this setup closely resembles models considered in the context of the island conjecture via a doubly holographic interpretation \cite{Almheiri:2019hni,Chen:2020uac}.

A plausible expectation might be that the intersection between the unbackreacted RT surface and the Nambu-Goto string, which on a constant–time slice appears as a point on the U-shaped string, defines the entangling surface from the worldsheet perspective. In this picture, the EE would measure the entanglement between the two segments of the string obtained by cutting the U at that point: one stretching from the left boundary endpoint to the intersection, and the other from the right endpoint to the intersection. In the symmetric configuration, this point coincides with the worldsheet horizon, which is consistent with the EE being identified with the black hole entropy in that case. If this interpretation of the intersection point continues to hold when $h\neq 0$, then varying the value of $h$ corresponds to shifting the location of the worldsheet ES.

Now that we understand a number of reasons to be interested in the asymmetric quark-antiquark setup, let us turn our attention to the question of whether  it is amenable to analytic computation.  In the gauge theory, the direct approach would be to use the replica trick, which would require evaluating the $h\neq 0$ Wilson loop in the replicated space. This is a highly non-trivial task, and to our knowledge, no such computation has been performed so far. Fortunately, on the gravity side, Ref.~\cite{Lewkowycz:2013nqa} provided prescription for computing the EE when the standard $U(1)$ symmetry that underpins the thermal case is broken. A priori this would require determining the string embedding on a backreacted replicated bulk geometry, which would again be a daunting task. But crucially, it was shown in \cite{Karch:2014ufa} that the leading-order result can be obtained without having to consider the backreaction. The calculation is still somewhat involved, but feasible. By following this approach, we will work out the EE contribution from the accelerated $q$-$\bar{q}$ pair, as a function of the distance parameter $h$ and the proper acceleration of the string $b^{-1}$, for the regime where $h \leq b$.\footnote{As explained in Appendix~\ref{Appendix A}, the case where  $h > b$ is more challenging and would require a more cumbersome treatment.} Conformal invariance dictates that the result be a function only of the dimensionless combination $h/b$.

The outline of the paper and summary of our results are as follows. In Section \ref{section 2}, we describe in more detail the system of interest, from both the CFT and AdS perspectives. As mentioned earlier, we are interested in a configuration where each particle is at a different distance from the ES. There are two equivalent ways of obtaining this setup from the usual equidistant configuration: we can laterally displace the ES, or the $q$-$\bar{q}$ pair. We adopt this second perspective, so that our coordinate system remains centered on the ES, which simplifies the passage to replica coordinates. In Euclidean signature, the trajectories of the $q$-$\bar{q}$ pair are then described  by a circle of radius $b$ and center at $(0,h)$ in the $x^0$-$x^1$ plane. The dual string embedding is obtained by solving the Nambu-Goto equation of motion with the $q$-$\bar{q}$ trajectories as boundary conditions. This embedding corresponds to an accelerating U-shaped string, that in Euclidean signature is described by a shifted hemispherical cap.

To compute the EE of this system, we use the replica trick, implemented in the gravitational theory. The replica space is well understood when the ES is a sphere. In subsection \ref{section 3.1}, we present the set of conformal transformations that map the ES, that in  the $x^{\mu}$ coordinates of \eqref{CFT Minkowski} is a plane at $x^0=x^1=0$, to a sphere described, in the new primed $x'^{\mu}$ coordinates, by $(x'^1)^2+(x'^2)^2+(x'^3)^2=b^2$. We also keep track of how the shifted circle describing the $q$-$\bar{q}$ trajectory changes under these transformations, and show that it is mapped to another circle of new radius $b'=\displaystyle{\frac{2b^3}{h(2b+h)}}$, centered at $\left(0,h'=\displaystyle{\frac{b(2b^2-h^2)}{h(2b+h)}}\right)$. In subsection~\ref{section 3.2} we then identify the equivalent transformations in the bulk, and examine how the string embedding changes. As expected, the initial shifted hemispherical cap is mapped to a new hemispherical cap with parameters $b'$ and $h'$. 

Before launching into the actual calculation, in Section~\ref{section 3.3} we discuss some features of the expected behavior of the quark-antiquark EE, using arguments from both sides of the duality. In particular, we argue that a divergence should occur in the limit $h/b \rightarrow 1$, where the quark intersects the ES. On the CFT side, we rely on the pointlike nature of the quark and on an argument based on the mutual information bound~\cite{Wolf:2007tdq}, while on the AdS side, we predict that the same divergence arises from the intersection between the RT surface and the Nambu–Goto string at the AdS boundary. Divergent contributions to the defect EE have also been encountered in the context of static, higher-dimensional defects, where intersection with the ES is unavoidable~\cite{Estes:2014hka,Kobayashi:2018lil}. In such cases, subtracting the EE of the ambient CFT, as prescribed in the definition  of the defect EE (for our system, this is the subtraction in (\ref{undisplacedentropy})), does not eliminate all divergences, even when a consistent UV cutoff is applied. This is because $S_{DCFT}$ contains not only the usual divergences associated with UV modes near the entangling surface, but also additional contributions localized on the defect itself, which generates extra divergences that are not present in the ambient CFT. In those scenarios, a renormalization procedure is necessary to isolate the universal part of the defect EE, which is physical and independent of the renormalization scheme. Notice that, in contrast to those cases, the divergence we encounter in our setup is intrinsically physical, arising from the universal contribution of the defect EE. 

As we will review in subsection \ref{section 4.1}, for quantum field theories that have a holographic description, the EE is obtained from the Euclidean gravitational action evaluated on the orbifold geometry $\hat{B}_n = B_n / \mathbb{Z}_n$. The concrete prescription \cite{Lewkowycz:2013nqa} is
$S = \left. \frac{\partial}{\partial n} \hat{I}[\hat{B}_n] \right|_{n=1}$,  with $B_n$ denoting the dual bulk solution whose boundary is the $n$-fold cover $\mathcal{M}_n$ appearing in the standard replica trick.  This framework can be extended to systems with flavor branes in the probe limit \cite{Karch:2014ufa}. As we will review in subsection \ref{HEE for probe branes}, the contribution of the flavor sector to the EE is computed directly from the on-shell action of the probe brane, eluding the need to compute the backreacted metric.  In subsection \ref{section 4.3}, we apply this technology to compute the EE contribution from the displaced uniformly accelerated $q$-$\bar{q}$ pair.  Working in hyperbolic coordinates (the natural adapted coordinates for the orbifold geometry $\hat{B}_n$) we show that the result decomposes into three distinct contributions: a contact term arising from the intersection of the string worldsheet with the Ryu-Takayanagi surface, a worldsheet integral of the string energy-momentum tensor and a counterterm contribution. This third term vanishes, consistent with the fact that the first two terms are finite. We thereby arrive at our main result, Eq.~(\ref{FinalresultEE}). We observe that this result reduces to \eqref{undisplacedentropy} in the limit $h/b \to 0$, as expected. Furthermore, in the limit $h/b \to 1$, our result displays a divergent behavior, in agreement with the prediction discussed in Section~\ref{section 3.3}. In Eq.~\eqref{FinalresultEEd}, we present a generalization of the defect entanglement entropy to the case of AdS$_{d+1}$/CFT$_d$ with arbitrary spacetime dimension $d$. Finally, we speculate on a possible interpretation of the two terms that contribute to the defect EE in $d=2$, drawing on an analogy with the structure of the generalized gravitational entropy in JT gravity~\cite{Almheiri:2019hni,Almheiri:2019psf}.

\section{The setup}\label{section 2}

\begin{figure}
  \centering
  \includegraphics[width=0.5\textwidth]{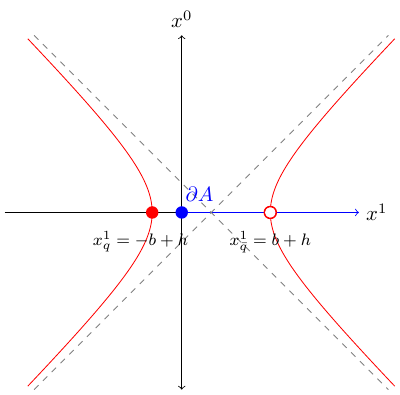}
  \caption{The $q$-$\bar{q}$  trajectories are shown in red. At $x^0=0$ the quark is at a distance $b-h$ and the antiquark at a distance $b+h$ from the entangling surface denoted as $\partial A$ (blue point).}
  \label{fig:0} 
  \end{figure}

We will start in the vacuum state of a CFT in $d=4$ Minkowski space. The metric is given by:
\begin{equation}
    ds^2=\eta_{\mu\nu}dx^{\mu}dx^{\nu}=-(dx^0)^2+(dx^1)^2+(dx^2)^2+(dx^3)^2.\label{CFT Minkowski}
\end{equation}
On the $x^0 = 0$ slice of this geometry, we will inquire into the EE  between a region $A$ defined as $x^1 > 0$ and its complement. The ES $\partial A$ is then the plane located at  $x^1 = 0$, as depicted in Fig.~\ref{fig:0}. We now introduce an infinitely massive quark and antiquark that separate back to back with uniform acceleration $b^{-1}$, following  hyperbolic trajectories. In this setup, we define a new geometric parameter $h\leq b$, which measures the distance between the ES and the center of the hyperbola.\footnote{For the nontrivial case where $h > b$, please refer to Appendix \ref{Appendix A}.} It is natural to expect that variations in $h$ will influence the EE, because they modify the bipartitioning of the gluonic field profile generated by the $q$-$\bar{q}$ sources, thereby changing the amount of entanglement between the two resulting subsystems.

 In the coordinates introduced above, the trajectories of the $q$ and $\bar{q}$ pair are given by
\begin{equation}
   (x^1-h)^2-(x^0)^2=b^2,\label{qq trajectories}
\end{equation}
where the center of the hyperbola is positioned at $(x^0,x^1)=(0,h)$. The trajectory of the quark is given by $x^1_{q}=-\sqrt{(x^0)^2+b^2}+h$, while that of the antiquark is  $x^1_{\bar{q}}=\sqrt{(x^0)^2+b^2}+h$. Initially, they travel towards each other from infinity, until they turn around at $x^0=0$ (with spatial locations at $x^1_{q}=-b+h$ for the quark and $x^1_{\bar{q}}=b+h$ for the antiquark) and then move away again to infinity. The quark and antiquark each occupy separate Rindler wedges, and they are never in causal contact throughout their motion.

The holographic dictionary tells us that the $q$-$\bar{q}$ pair moving with constant acceleration is dual to an accelerating U-shaped open string in pure AdS, with the two endpoints attached to a flavor brane \cite{Maldacena:1998im,Rey:1998ik,Xiao:2008nr}. In Type IIB string theory, this would be a $D7$-brane\footnote{More precisely, the endpoints of the string are dual to the $q$-$\bar{q}$ pair, and the body of the string is dual to the color flux tube that joins the pair (due to the non-confining nature of the theory, this flux `tube' is actually spread out, and even accounts for gluonic radiation) \cite{Danielsson:1998wt,Callan:1999ki,Garcia:2012gw,Chernicoff:2011xv}.}. In Poincaré coordinates,
\begin{equation}
ds^2=\frac{L^2}{z^2}(-(dx^0)^2+(dx^1)^2+(dx^2)^2+(dx^3)^2+(dz^2)),
\end{equation}
the position of this brane is inversely proportional to the mass of the quark, $z_{m}=\frac{\sqrt{\lambda}}{2 \pi m}$, so for an infinitely massive pair, the endpoints of the string would be attached to the boundary at $z=0$. From now on, we will refer only to this scenario.

The dynamics of the string is characterized by the Nambu-Goto action,
\begin{equation}
    I_{NG}=-T \int{d^2\sigma} \sqrt{-\gamma},\label{NG action}
\end{equation} 
where $T$  is the tension of the string and $(\sigma^0,\sigma^1)$ are the string worldsheet coordinates. The induced metric is $\gamma_{ab}=\partial_aX^{\mu}\partial_bX_{\mu}$, with  $a,b= 0$ or $1$.  We can work in the static gauge, taking $(\sigma^0,\sigma^1)$ to be $(x^0,z)$ and $X^{\mu}=(x^0,z,x^1(x^0,z),0,0)$. In this gauge, the equations of motion are
\begin{equation}
    \frac{\partial }{\partial z}\left(\frac{\partial_z x^1}{z^{2}\sqrt{1+(\partial_z x^1)^2-(\partial_0 x^1)^2}}\right)-\frac{\partial }{\partial x^0}\left(\frac{\partial_0 x^1}{z^{2}\sqrt{1+(\partial_z x^1)^2-(\partial_0 x^1)^2}}\right)=0.\label{eomNambugoto}
\end{equation}

In Ref. \cite{Xiao:2008nr}, a solution to the Nambu-Goto equations for a uniformly accelerated string was presented. Here, we introduce a shifted version of this solution, which of course remains a valid solution to the equation of motion due to the invariance  of  \eqref{eomNambugoto} under translations in $x^1$. The embedding is described by the equation:
\begin{equation}
    (x^1-h)^2-(x^0)^2+z^2=b^2,\label{string in AdS}
\end{equation}

The endpoints of the string, located at $z=0$, correspond to the positions $x^1_{q}$ and $x^1_{\bar{q}}$ mentioned before. The induced metric is:
\begin{equation}
    ds^2_{\gamma}=\frac{L^2}{z^2\left(b^2-z^2+(x^0)^2\right)}\left[-(b^2-z^2)(dx^0)^2-2zx^0dx^0dz+(b^2+(x^0)^2)dz^2\right].
\end{equation}

One of the most important aspects of this worldsheet metric is that it presents two horizons at $z=b$ and exhibits a causal structure similar to that of an eternal AdS black hole. This is why this configuration has been a subject of study in the context of the ER=EPR conjecture \cite{Maldacena:2013xja}, as it provides a realization of it \cite{Jensen:2013ora,Sonner:2013mba,Chernicoff:2013iga}.

In the next section, we will perform a set of conformal transformations in the CFT (corresponding to a set of diffeomorphisms in the bulk), which will translate the problem into a more suitable one for studying the EE in this configuration.

\section{Coordinate transformations}\label{section 3}
    
\subsection{Coordinate transformations in the CFT}\label{section 3.1}

\begin{figure}
 \centering
  \includegraphics[width=1\textwidth]{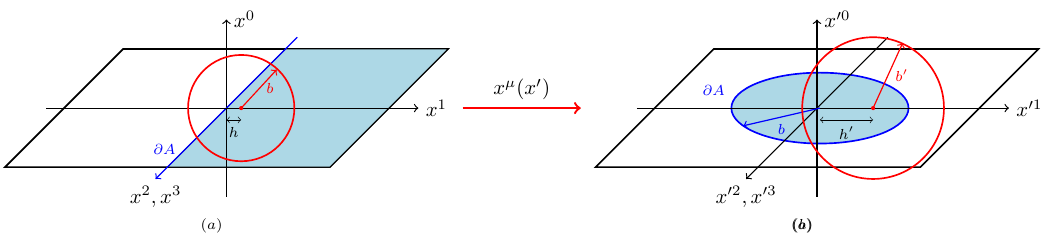}
  \caption{(a)  These are the trajectories of the   $q$-$\bar{q}$ pair after a Wick rotation. They are described by a shifted circle with its origin located at $(0,h)$. (b) This is the result of applying a set of conformal transformations to the configuration in (a). The ES is mapped from a hyperplane to a sphere (blue circle) of radius $r'=b$ and the $q$-$\bar{q}$ trajectory  is mapped to a shifted circle in the $x'^0$-$x'^1$ plane with a new radius $b'$ (red lines).}
  \label{fig:conformal T} 
  \end{figure}

Since we are interested in the EE, we will work in Euclidean signature. Therefore, we perform a Wick rotation on \eqref{CFT Minkowski},  $x^0 \rightarrow -ix^0$, obtaining that the  $q$-$\bar{q}$ pair trajectories are described by a circle of radius $b$  shifted by a distance $h$ from the origin of coordinates, as shown in Fig.~\ref{fig:conformal T}a. In this picture, the left half of the circle corresponds to the quark, and the right half to the antiquark. Following \cite{Lewkowycz:2013laa}, we then transition to double polar coordinates using the following transformations:
\begin{align}
   x^0&=\text{r} \sin(\tau), \label{X0 DP}\\
   x^1&=\text{r} \cos(\tau),\label{X1 DP}\\
    x^2&=y \cos(\phi),\label{X2 DP}\\ 
    x^3&= y \sin(\phi) ,\label{X3 DP}
\end{align}
where $r\geq 0$ and $y\geq 0$ are radial coordinates, and $0<\tau,\phi<2\pi$. The resulting metric is:
 \begin{equation}
ds^2=\text{r}^2d\tau^2+d\text{r}^2+dy^2+y^2d\phi^2.\label{double polars}
 \end{equation}
In these new coordinates,  the entanglement region $A$ is situated at $\text{r}\geq 0$ and $\tau=0$, placing the ES precisely at $\text{r}=0$. These coordinates will serve as the replica coordinates discussed in section \ref{HEE for probe branes}.
 
After the change of signature and coordinates, the  $q$-$\bar{q}$ trajectories are  described by the following parametric equation:
\begin{align}
    \text{r}(\tau)&=h \cos(\tau) + \sqrt{b^2-h^2 \sin^2(\tau)} \:, \label{embedding r cft}\\
    y&=0\label{embedding y cft}.
\end{align}
Notice that this reduces to $r=b$ when $h=0$, as one would expect.
 
We will perform two conformal transformations to map the hyperplane to a sphere. First, we will switch to hyperbolic coordinates, since they will be useful when obtaining the associated diffeomorphism in the dual bulk space. These transformations are:
\begin{align}
 \cosh(\rho)=\frac{b^2+y^2+\text{r}^2}{2\text{r}b}\:,\label{Hyperbolic rho}\\
 \cot(\theta)=\frac{b^2-y^2-\text{r}^2}{2yb}   \:.\label{Hyperbolic theta}
\end{align}
The metric is then transformed as:
\begin{equation}
ds^2=\text{r}^2ds^2_{S^1\times H^{3}} \:\text{, where}\: \:ds^2_{S^1\times H^{3}}=d\tau^2+d\rho^2+\sinh(\rho)^2\left(d\theta^2+\sin(\theta)^2d\phi^2\right) \:.   
\end{equation}

In these coordinates the ES is located at $\rho=\pm \infty$ and along $\theta$ and $\phi$. Now, we perform the second conformal transformation
\begin{align}
t'&=b\frac{\sin(\tau)}{\cosh(\rho)+\cos(\tau)}\:, \label{CFT prime t}\\
r'&=b\frac{\sinh(\rho)}{\cosh(\rho)+\cos(\tau)} \: ,\label{CFT prime r}
\end{align}
obtaining the metric $ds'^2=(\frac{\cosh(\rho)+\cos(\tau)}{b})^{-2}ds^2_{S^1\times H^{3}}$ where  $ds'^2$ denotes Minkowski space in spherical coordinates:
\begin{equation}
ds'^2=dt'^2+dr'^2+r'^2(d\theta^2+\sin(\theta)^2d\phi^2)\:.
\end{equation}
In this coordinate system, the ES forms a sphere described by $r'=b$ at $t'=0$. 

Tracing this sequence of conformal transformations, it is easy to see  that the map relating the primed  space to the double polar space is
\begin{align}
     t'&=\frac{2\text{r}b^2\sin(\tau)}{b^2+y^2+\text{r}^2+2\text{r}b\cos(\tau)}~, \label{t' transformation}\\
      r'&=b\frac{\sqrt{(b^2+y^2+\text{r}^2)^2-(2\text{r}b)^2}}{b^2+y^2+\text{r}^2+2\text{r}b\cos(\tau)}~, \label{r' transformation}\\
      \theta&=\cot^{-1}\left(\frac{b^2-y^2-\text{r}^2}{2yb}\right).\label{theta transformation}
\end{align}
   
The corresponding metrics are related by $ds'^2=\displaystyle{\left[\frac{2b^2}{b^2+y^2+\text{r}^2+2\text{r}b\cos(\tau)}\right]^2ds^2}$. To obtain the $q$-$\bar{q}$ pair trajectories in these primed coordinates, we need to substitute  \eqref{embedding r cft}-\eqref{embedding y cft} into  \eqref{t' transformation}-\eqref{r' transformation}, resulting in
\begin{align}
 t'(\tau)&=b\frac{2b\left(h\cos(\tau)+\sqrt{b^2-h^2\sin(\tau)^2}\right)\sin(\tau)}{2b^2-h^2+2(h+b)\left(h\cos(\tau)+\sqrt{b^2-h^2\sin(\tau)^2}\right)\cos(\tau)}~,\label{t' embedding}\\
       r'(\tau)&=b\frac{\left|h^2-2h^2\cos(\tau)^2-2h\cos(\tau)\sqrt{b^2-h^2\sin(\tau)^2}\right|}{2b^2-h^2+2(h+b)\left(h\cos(\tau)+\sqrt{b^2-h^2\sin(\tau)^2}\right)\cos(\tau)}~,\label{r' embedding}\\  
       \theta&=0  \; \text{and} \;\theta=\pi~.\label{theta embedding}
\end{align}
 It is easy to see that in the case where $h=0$, the trajectory in the primed coordinates is described by a straight line located at $r'=0$, and running along the $t'$ coordinate.   

We can rewrite these expressions in Cartesian coordinates using the usual transformations:  $x'^0=t'$, $x^1=r'\cos{\theta}$, $x'^2=r'\sin{\theta}\cos{\phi}$ and $x'^3=r'\sin{\theta}\sin{\phi}$.
From \eqref{theta transformation} and \eqref{embedding r cft}-\eqref{embedding y cft}, we can infer that the trajectories of the $q$-$\bar{q}$ pair are located at $\theta=0$ when $b^2-\text{r}(\tau)^2\geq 0$ and at $\theta=\pi$ when $b^2-\text{r}(\tau)^2\leq 0$. Now, plugging \eqref{embedding r cft}-\eqref{embedding y cft} into  \eqref{r' transformation},  we  see that the denominator of  the latter equation becomes  $|b^2-\text{r}(\tau)^2|$. Due to this and the fact that $x'^1(\tau)=r'(\tau)$ for $\theta=0$ and $x'^1(\tau)=-r'(\tau)$ for $\theta=\pi$, the embedding in Cartesian coordinates will appear as
\begin{align}
       x'^0(\tau)&=b\frac{2b(h\cos(\tau)+\sqrt{b^2-h^2\sin(\tau)^2})\sin(\tau)}{2b^2-h^2+2(h+b)(h\cos(\tau)+\sqrt{b^2-h^2\sin(\tau)^2})\cos(\tau)}~,
       \\
x'^1(\tau)&=b\frac{h^2-2h^2\cos(\tau)^2-2h\cos(\tau)\sqrt{b^2-h^2\sin(\tau)^2}}{2b^2-h^2+2(h+b)(h\cos(\tau)+\sqrt{b^2-h^2\sin(\tau)^2})\cos(\tau)}~,
\\  
       x'^2&=0  \; \text{and} \; x'^3=0~.
\end{align}

It is straightforward to see that this embedding satisfies the relation
\begin{equation}
    (x'^1-h')^2+(x'^0)^2=b'^2~, \label{prime circle}
\end{equation}
where 
\begin{equation}
    h'=\displaystyle{\frac{b(2b^2-h^2)}{h(2b+h)}}\quad,\quad b'=\displaystyle{\frac{2b^3}{h(2b+h)}}~.\label{transform parameters}
\end{equation}
 As illustrated in Fig.~\ref{fig:conformal T}b, this represents a shifted circle in the $x'^0$-$x'^1$ plane. This result is expected, as all we have done is apply conformal transformations, which can only translate and rescale the shifted circle.

\begin{figure}
  \centering
  \includegraphics[width=0.5\textwidth]{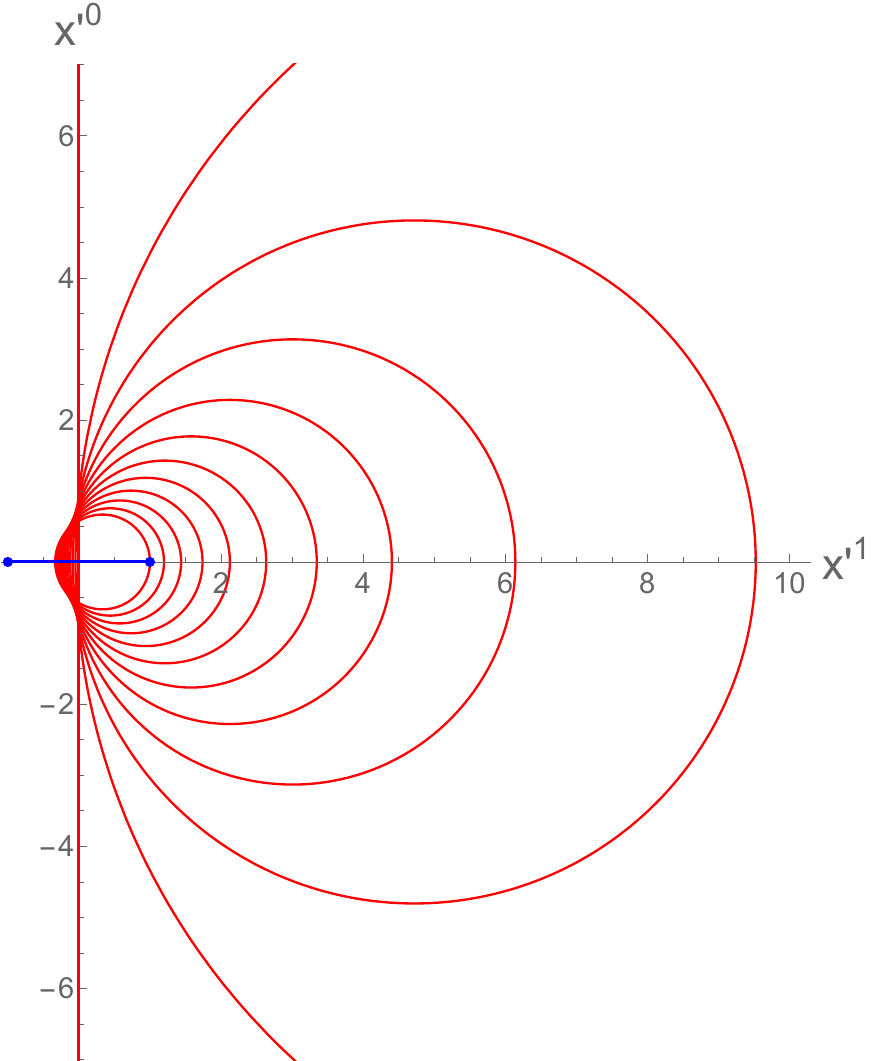}
  \caption{The red lines represent the $q$-$\bar{q}$ trajectories in the primed coordinates. We have plotted the trajectories for different values of $h$ (taking $b=1$). The blue points indicate the locations of the spherical ES. The $h=0$ plot corresponds to the vertical line, while  the  $h=1$ plot corresponds to the circumference that intersects the ES.}
  \label{fig:3} 
  \end{figure}

In Fig.~\ref{fig:3}, we have plotted the trajectories in the $x'^0$-$x'^1$ plane for an initial radius $b = 1$ and various values of $h \leq b = 1$. As observed, when $h$ approaches 0, the values of $h'$ and $b'$ increase significantly. In the limit $h \rightarrow 0$, the trajectories become a vertical Wilson line located at the center of the entangling region, consistent with expectations \cite{Lewkowycz:2013laa}. Conversely, as $h$ approaches $b$, the radius $b'$ decreases, and when $h = b$, the quark trajectory intersects the ES.

\subsection{Coordinate transformations in AdS}\label{section 3.2}

 In Euclidean signature, the string embedding of \eqref{string in AdS} becomes a shifted spherical cap,
\begin{equation}
    (x^1-h)^2+(x^0)^2+z^2=b^2,
\end{equation}
where, as mentioned, $b^{-1}$ is the proper acceleration of the string. 

We  pass to a new type of coordinates, which we will call AdS double polar coordinates, by the transformation
\begin{align}
    z&=r/u, \label{Bulk DP1}\\
    x^0&=\sqrt{1-1/u^2}\,r\sin(\tau)~,\label{Bulk DP2}\\
     x^1&=\sqrt{1-1/u^2}\,r\cos(\tau)~,\label{Bulk DP3}
\end{align}
together with \eqref{X2 DP}-\eqref{X3 DP}, where  $1<u<\infty$. It is important to notice that this $r$ coordinate is not the same as the r appearing in  \eqref{X0 DP}-\eqref{X1 DP}, because this coordinate goes deep into the bulk. However,  when we approach the AdS boundary by taking the asymptotic limit $u \rightarrow \infty$, \eqref{Bulk DP2}-\eqref{Bulk DP3}  become  \eqref{X0 DP}-\eqref{X1 DP},  and therefore $r\rightarrow \text{r}$. 

In these  hyperbolic coordinates, the equation describing the accelerated string is parameterized by $\tau$ and $u$ as follows:
\begin{align}
    r(\tau,u)&=h\sqrt{1-1/u^2} \cos(\tau) + \sqrt{b^2-h^2+ h^2(1-1/u^2)\cos(\tau)^2}~,
    \label{string dp r} \\
    y&=0~. \label{string dp y}
\end{align}
Of course, when we go to the boundary we recover the $q$-$\bar{q}$ trajectories. 

Now we use the transformations of \eqref{Hyperbolic rho}-\eqref{Hyperbolic theta} (with the subtlety mentioned before for the $r$ coordinate) to obtain an $S^1 \times H^3$ foliation of Euclidean AdS$_5$ space. The metric in these coordinates will be
\begin{equation}
    ds^2=L^2\left[(u^2-1)d\tau^2+\frac{du^2}{u^2-1}+u^2(d\rho^2+\sinh(\rho)^2\left(d\theta^2+\sin(\theta)^2d\phi^2)\right)\right]~.\label{hyperbolic AdS metric}
\end{equation}
An important feature of this metric is that, as noticed in \cite{Myers:2010xs,Myers:2010tj}, it can be interpreted as a topological black hole \cite{Emparan:1999gf}  with the horizon located at $u_h=1$.

Finally, we can return to a primed version of the  Poincaré patch by the transformations
\allowdisplaybreaks
\begin{align}
z'&=b\frac{1}{u\cosh(\rho)+\sqrt{1-u^2}},\label{z' bulk transformation}\\
  t'&=b\frac{\sqrt{1-u^2}\sin(\tau)}{u\cosh(\rho)+\sqrt{1-u^2}\cos(\tau)}, \label{t' bulk transformation}\\
r'&=b\frac{u\sinh(\rho)}{u\cosh(\rho)+\sqrt{1-u^2}\cos(\tau)} .\label{r' bulk transformation}
\end{align}
This yields the metric
\begin{equation}
ds'^2=\frac{L^2}{z'^2}\left[dz'^2+(dt')^2+r'^2(d\theta^2+\sin(\theta)^2d\phi^2)\right]~.
\end{equation}
Again, the transformations \eqref{z' bulk transformation}-\eqref{r' bulk transformation} are just the bulk diffeomorphism that corresponds  to the conformal boundary transformations of \eqref{t' transformation}-\eqref{r' transformation}. 

The whole point of performing these transformations is to map the ES from a plane to a sphere, which is a useful starting point for EE computations \cite{Myers:2010tj,Myers:2010xs,Casini:2011kv}. In this case, the Ryu-Takayanagi (RT) surface will be a hemisphere hanging from the ES at the boundary.

Returning to our discussion, if we keep track of the transformations that we have made, we can relate the primed Poincaré coordinates to the bulk  double polar coordinates as follows
\begin{align}
z'&=\frac{2rb^2}{u(b^2+y^2+r^2+2rb\sqrt{1-1/u^2}\cos(\tau))}~,
\label{z' to dp}\: \\
  t'&=\frac{2rb^2\sqrt{1-1/u^2}\sin(\tau)}{b^2+y^2+r^2+2rb\sqrt{1-1/u^2}\cos(\tau)}~, 
  \label{t' to dp}\\
r'&=\frac{b\sqrt{(b^2+y^2+r^2)^2-(2rb)^2}}{b^2+y^2+r^2+2rb\sqrt{1-1/u^2}\cos(\tau)}~,
\label{r' to dp}\\
\theta&=\cot^{-1}\left(\frac{b^2-y^2-r^2}{2yb}\right)~.\label{theta to dp}
\end{align}

To obtain the string embedding in these new coordinates, we need to substitute \eqref{string dp r}-\eqref{string dp y} into \eqref{z' to dp}-\eqref{theta to dp}. We obtain
\footnotesize\begin{align}
    z'&=\frac{2b^2\left(h\sqrt{1-1/u^2} \cos(\tau) + \sqrt{b^2-h^2+ h^2(1-1/u^2)\cos(\tau)^2}\right)}{u\left(2b^2-h^2+2(b+h)\cos(\tau)\left(h(1-1/u^2)\cos(\tau)+\sqrt{(1-1/u^2)(b^2-h^2+h^2(1-1/u^2)\cos(\tau)^2)}\right)\right)}\\
    t'&=\frac{2b^2\left(h(1-1/u^2) \cos(\tau) +\sqrt{(1-1/u^2)(b^2-h^2+h^2(1-1/u^2)\cos(\tau)^2)}\right)\sin(\tau)}{2b^2-h^2+2(b+h)\cos(\tau)\left(h(1-1/u^2)\cos(\tau)+\sqrt{(1-1/u^2)(b^2-h^2+h^2(1-1/u^2)\cos(\tau)^2)}\right)}\\
r'&=\frac{b\left|h^2-2h^2(1-1/u^2)\cos^2(\tau)-2h\cos(\tau)\sqrt{(1-1/u^2)(b^2-h^2+h^2(1-1/u^2)\cos(\tau)^2)}\right|}{2b^2-h^2+2(b+h)\cos(\tau)\left(h(1-1/u^2)\cos(\tau)+\sqrt{(1-1/u^2)(b^2-h^2+h^2(1-1/u^2)\cos(\tau)^2)}\right)}\\
 \theta&=0  \: \text{and} \:\theta=\pi.
\end{align}
\normalsize
We can rewrite the expressions in Cartesian coordinates following a procedure analogous to the one of the previous section. Doing so, it is easy to notice that the string embedding in these coordinates is also a spherical cap, described by the equation
\begin{equation}
    (x'^1-h')^2+(x'^0)^2+z'^2=b'^2,
\end{equation} 
which of course reduces to \eqref{prime circle} at the boundary $z'=0$.

\section{Expected qualitative features of the EE}\label{section 3.3}

Before performing the explicit holographic calculation in Section \ref{section 4}, it is instructive to discuss some qualitative expectations for the EE of the displaced $q$-$\bar{q}$ pair. Conformal invariance dictates that the result can depend only on the dimensionless ratio $\frac{h}{b}$, and we can anticipate that the EE will interpolate smoothly between the symmetric case $\frac{h}{b}\to 0$ given in \eqref{undisplacedentropy} and the configuration $\frac{h}{b}\to 1$, where the quark approaches the ES.

From the CFT perspective, as discussed in Section \ref{section 3.1}, the setup can be conformally mapped to a geometry where the entangling surface is a sphere of radius $b$. In this picture, the trajectory of the $q$-$\bar{q}$ pair is mapped to a circle in the $x^{\prime 0}$-$x^{\prime 1}$ plane with radius $b'$ and shift $h'$, related to the original variables by \eqref{transform parameters}. This representation makes it easier to visualize the behavior of what one might loosely call a defect entanglement entropy, even though our defect is non-static and does not preserve a subgroup of the conformal group as in standard DCFTs. In this image, the symmetric limit $\frac{h}{b}\to 0$ corresponds to $b'\to\infty$ and/or $h'\to 0$, reproducing the familiar Wilson line configuration, with the quark infinitely far away. As $\frac{h}{b}$ increases, the quark moves closer to the entangling surface (and to the antiquark). See Fig.~\ref{fig:3}. The quark acts as a pointlike source that excites higher UV modes of the gluonic field as one approaches it. It is natural to expect that these modes contribute to the EE, increasing its value as the distance between the quark and the ES decreases, which will ultimately lead to divergent behavior in the limit where this distance approaches zero. In other words, the issue is that the complete cancelation of UV divergences due to the subtraction in (\ref{undisplacedentropy}) only applies when both terms share the same vacuum structure at the ES. This ceases to be true precisely when the ES overlaps with the pointlike defect.

The divergent behavior can also be anticipated from the mutual information  bound for quantum correlators \cite{Wolf:2007tdq},
\begin{equation}
    I(A,B) \geq \frac{ (\langle O_A O_B \rangle - \langle O_A \rangle \langle O_B \rangle )^2}{2\|O_A\|^2 \|O_B\|^2}\:,\label{correlators bound}
\end{equation} 
where $ I(A,B) =S(A)+S(B)-S(A\cup B)$ is the mutual information and $\|O_A\|^2$ and $\|O_B\|^2$ are the absolute values of the maximum eigenvalue of the operators $O_A$ and $O_B$ respectively. This inequality is well defined when $O_A$ and $O_B$  are bounded operators.  In our case of interest, the correlation functions and mutual information are computed in the presence of a quark-antiquark pair, whose Euclidean trajectory can be represented as the insertion of a circular Wilson loop.

If we take the $A$ and $B$ regions in  \eqref{correlators bound} to be well separated, then
the leading-order mutual information  vanishes at order $\mathcal{O}(N^2)$. This follows because $S(A\cup B)=S(A)+S(B)$ at this order, which is intuitive from the holographic picture, as the EE is entirely determined by disjoint RT surfaces. However, at subleading order $\mathcal{O}(N^0)$, the defect produces a nonzero contribution, which is $\mathcal{O}(\sqrt{\lambda})$ on account of the Wilson loop. This contribution is necessary to obtain nonvanishing correlators \cite{VanRaamsdonk:2009ar,VanRaamsdonk:2010pw,Faulkner:2013ana}.

  \begin{figure}
  \centering
\includegraphics[width=0.4\textwidth]{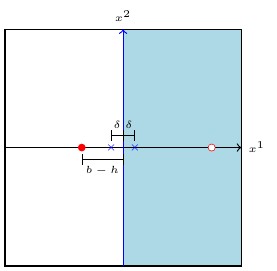}
  \caption{The figure shows the $x^1$–$x^2$ plane at time $t=0$. Region $A$ (shaded in light blue) is defined by $x^1 \geq 0$, while region $B$ is its complement. The entanglement surface lies along $x^1 = 0$ (indicated by the blue line). The bounded  operators $O_A$ and $O_B$ are inserted at $x^1 = \delta$ and $x^1 = -\delta$, respectively, and are represented by crosses ($\times$). The Wilson loop intersects the $t=0$ plane at $x^1_q = -b + h$ (red circle) and $x^1_{\bar{q}} = b + h$ (white circle).}
  \label{fig:6} 
  \end{figure}

Let us return now to our configuration of interest, in the original, unprimed coordinates. We are interested in the case where $B$ is the complement of $A$ and the overall state is pure. This implies that
 $I(A,B) =2S(A)$, and therefore, at order $\mathcal{O}(N^0\sqrt{\lambda})$, the left hand side of \eqref{correlators bound} will be determined completely by the defect entropy. The bound \eqref{correlators bound} will clearly be most restrictive when the operator in region $A$  is as close as possible to the operator in region $B$. So, as depicted in Fig.~\ref{fig:6}, we place bounded operators $O_A$ and $O_B$ at a fixed but small distance $\delta$ from the ES. On account of their proximity, we can use the operator product expansion (OPE) to replace them by a sum over local operators placed at the midpoint between them, right on the ES. The desired two-point correlator would normally be dominated by the one-point function of the Wilson line with the operator in the OPE that has the lowest conformal dimension.  This one-point function probing the gluonic field is naturally expected to increase and become divergent as the insertion approaches the quark.\interfootnotelinepenalty=10000\footnote{In more detail, the different scales involved in this description need to be appropriately synchronized. If we denote with $\eta$ the size of the compact support of the operators $O_A,O_B$, with $\delta$ the inter-operator distance, and with $\epsilon\equiv b-h$ their distance to the quark, we would like to have  $\eta\ll \delta \ll \epsilon$. This is subtle because $\eta$ controls the operator norms in the denominator of (\ref{correlators bound}), and $\epsilon$ determines the radius of convergence of the OPE, implying that the putative divergence is not necessarily controlled by the operator with the lowest conformal dimension. }  This has been examined for one-point and two-point functions of certain local operators in \cite{Buchbinder:2012vr,Barrat:2020vch, Billo:2024kri}. In our context, this corresponds to the limit $h\rightarrow b$.  Consequently, the left-hand side of \eqref{correlators bound} would also have to diverge at this order. Given that the mutual information in our case is twice the defect entropy, the latter would then be predicted to also diverge at order $\mathcal{O}(N^0\sqrt{\lambda})$.
 Phrasing this in the opposite direction, through (\ref{correlators bound}) we learn that a divergent defect entropy would leave margin for divergent two-point correlators of bounded operators.

We can provide a complementary heuristic 
argument for a divergent behavior by anticipating one expected feature of the result from the bulk computation (which we will carry out in detail in the next section). First, note that increasing $\frac{h}{b}$ corresponds to moving the intersection point between the string worldsheet and the RT surface closer to the AdS boundary. As discussed in Section~\ref{introduction}, this point is naturally expected to play an important role in the holographic EE computation, perhaps even serving as a worldsheet ES. Given the asymptotic behavior of the induced metric on the string, it is thus plausible to expect the EE to grow and eventually diverge as $\frac{h}{b}\to 1$. In fact, through the UV/IR correspondence \cite{Susskind:1998dq,Peet:1998wn,Agon:2014rda}, the fact that the string approaches the boundary at the location of the quark is precisely dual to our statement about gluonic UV modes three paragraphs above.

\section{Holographic Entanglement Entropy for the shifted configuration}\label{section 4}

\subsection{Generalized gravitational entropy method}\label{section 4.1}

In the context of the AdS/CFT correspondence, a fundamental question concerns the interpretation of the EE on the gravitational side of the duality. In Refs \cite{Ryu:2006bv,Ryu:2006ef} it was conjectured that the EE is proportional to a minimum area of a codimension-two surface $\gamma_A$ in the bulk (the interior of AdS) homologous to the boundary region $A$ of interest,

\begin{equation}
    S_A=\frac{\mathrm{Area}(\gamma_A)}{4G_N} \label{RT formula}.
\end{equation}

In this section we will review the derivation of the RT formula for computing EE \cite{Lewkowycz:2013nqa}. Let us start by mentioning some results concerning the EE of a spherical entangling region. In Ref.\cite{Casini:2011kv} it was shown that, by making a sequence of conformal transformations, it is possible to relate the vacuum state of the original geometry to a thermal state placed on an $R\times H^{d-1}$ background \eqref{CFT prime t}-\eqref{CFT prime r}. Then, the AdS/CFT dictionary was used to translate this problem to the one of finding the horizon area entropy of a certain topological black hole. Furthermore, this procedure was extended for a CFT with defects where it was shown that for a defect centered with respect to the ES, the same thermal interpretation can be given \cite{Jensen:2013lxa}. This case is crucial for our discussion, as the $q$-$\bar{q}$ pair can be thought as a codimension $d-1$ defect on the CFT. Since our discussion encompasses more than just the scenario where the defect is centered on the ES (which corresponds to the case $h = 0$), we will focus on a proof of \eqref{RT formula} that does not rely on this thermal interpretation.

The method builds upon a generalization of the Euclidean path integral approach to gravitational entropy \cite{Gibbons:1976ue}, extending it to cases without a $U(1)$ symmetry \cite{Lewkowycz:2013nqa}. In the framework of AdS/CFT correspondence, this generalized gravitational entropy \footnote{Not to be confused with Bekenstein's notion of generalized entropy.} is dual to the von Neumann entropy of a density matrix on the CFT side.

The extension of gravitational entropy beyond $U(1)$ symmetry was achieved by the generalization of the conventional replica trick, which is commonly utilized in QFT, to the bulk. From the field theory standpoint, the replica method is especially valuable for computing Rényi entropy, which can be depicted in terms of the partition function evaluated on an $n$-fold cover $\mathcal{M}_n$ of the original $d$-dimensional space $\mathcal{M}$. This manifold is derived by cyclically gluing $n$ copies along the entangling region of interest, exhibiting a manifestly $\mathbb{Z}_n$ symmetry \cite{Calabrese:2004eu,Holzhey:1994we}.  Rényi entropy is defined as follows:

\begin{equation}
    S_{n}=\frac{1}{1-n}\ln{\left(\frac{Z[\mathcal{M}_n]}{Z[\mathcal{M}]^n}\right)}. \label{Renyi Entropy}
\end{equation}

In the holographic picture, the dictionary suggests that a bulk space $\mathcal{B}_n$ is dual to the boundary $\mathcal{M}_n$. The $\mathbb{Z}_n$ symmetry of the boundary is usually inherited by the bulk. Specifically, the operation of the $\mathbb{Z}_n$ symmetry allows us to define the bulk orbifold $\hat{\mathcal{B}}_n=\mathcal{B}_n/\mathbb{Z}_n$, which is regular everywhere except at fixed points. At these points, a conical singularity appears with a deficit angle of $2\pi(1-1/n)$. Utilizing the Euclidean version of the GKPW relation \cite{Gubser:1998bc,Witten:1998qj} in the saddle point approximation, the Renyi entropy would be given by
\begin{equation}
S_n=\frac{n}{n-1}(\hat{I}[\hat{\mathcal{B}}_n]-I[\mathcal{B}_1]),
\end{equation}
where $\hat{I}[\hat{\mathcal{B}}_n]$ is the bulk-per-replica action, defined as $I[\mathcal{B}_n]/n$. It is crucial to differentiate it from the action of the quotient space, $I[\hat{\mathcal{B}}_n]$, which has  a contribution coming from the conical singularity located along the codimension-two hypersurface $\gamma_A^{(n)}$. For the case of Einstein-AdS the  bulk-per-replica action is
\begin{equation}
   \hat{I}[\hat{\mathcal{B}}_n]=I[\hat{\mathcal{B}}_n]+\frac{1}{4G}\left(1-\frac{1}{n}\right)A[\gamma^{(n)}].\label{Action in replica orbifold}
\end{equation}

The EE is then obtained from the $n=1$ limit of the aforementioned equation, yielding
\begin{equation}
    S= \lim\limits_{n\rightarrow 1} S_n=\partial_n\hat{I}[\hat{\mathcal{B}}_n]|_{n=1}, \label{LM EE}
\end{equation}
where the partial derivative with respect to $n$ appears because we are using L'H\^opital's rule when taking the limit. Replacing \eqref{Action in replica orbifold} in \eqref{LM EE} and  noticing that $\partial_n \hat{I}[\hat{\mathcal{B}}_n]=\frac{\partial g_{\mu \nu}}{\partial n}\frac{\delta\hat{I}[\hat{\mathcal{B}}_n]}{\delta g_{\mu \nu}}+\frac{A[\gamma^{(n)}]}{4Gn^2}$ and $\frac{\delta\hat{I}[\hat{\mathcal{B}}_n]}{\delta g_{\mu \nu}}=0$ (because $\hat{I}[\hat{\mathcal{B}}]$ satisfies the equation of motion), then
\begin{equation}
    S=\frac{A_{\text{min}}[\gamma]}{4G}.
\end{equation}

One can argue that the minimality of the surface  $\gamma$ is guaranteed in this derivation. This can be seen heuristically using the cosmic brane interpretation \cite{Lewkowycz:2013nqa,Dong:2016fnf}. When one takes the limit $n \rightarrow 1$, the tension of the cosmic brane vanishes, so there is no backreaction on the background bulk geometry $\hat{\mathcal{B}}_{n}$. This means that the position of the brane must satisfy $\delta A_1=0$,  thus ensuring the minimality condition for  $\gamma$.

\subsection{EE for Probe Branes}\label{HEE for probe branes}

The string attached to a $q$-$\bar{q}$ pair at the boundary can be conceptualized as a probe brane with $p=1$ dimensionality. In typical scenarios, the probe limit implies that there is no  backreaction coming from the dynamics of the brane. However, in the context we are discussing, when we refer to the probe limit, what we are actually indicating is that the first-order contribution to the EE (denoted as $\mathcal{O}(t)$, with $t$ being the tensional parameter controlling the backreaction)  can be computed using the original metric background without taking into account the backreaction.

In this section we will reproduce the results obtained in \cite{Karch:2014ufa} for the EE  contribution of probe $p$-branes using the generalized gravitational entropy method described in the previous section. This problem was addressed before in \cite{Chang:2013mca,Jensen:2013lxa}, but those prior computations relied heavily on a spherical ES with the probe brane located at the center, which is different from our case of interest.

The Euclidean action of interest is
\begin{equation}
    I=I_{\mbox{\tiny EH}}+I_{\mbox{\tiny NG}}~.
    \label{bulk plus brane action}
\end{equation}

In this discussion, we will use coordinates adapted  to the RT surface $\gamma$, that describe the near-conical-singularity geometry mentioned before. In particular, the $(\tau, r)$ coordinates will represent the two-dimensional orthogonal space to the RT surface. The $\tau$ coordinate is the $S_1$ direction that implements the $\mathbb{Z}_n$ replica symmetry mentioned before (where $n$ can be a non-integer number) and the locus  where the $S_1$  degenerates is located at $r=0$ . Since we explicitly focus on aAdS spaces, both bulk and brane actions require renormalization. This would be done by putting a cutoff at $r=r_{\epsilon}$ (with $r_{\epsilon}\rightarrow \infty$ at the end), and including the corresponding counterterms. 

Breaking down \eqref{bulk plus brane action}, we have
\begin{align}
    I_{\mbox{\tiny EH}}&=\int dr d^dx\mathcal{L}_{\mbox{\tiny EH}}+ I_{\mbox{\tiny EH}}^{\mbox{\tiny ct}},\\
      I_{\mbox{\tiny NG}}&=\int dr d^py\mathcal{L}_{\mbox{\tiny NG}}+ I_{\mbox{\tiny NG}}^{\mbox{\tiny ct}}.
\end{align}

We will be interested in $\mathcal{L}_{\mbox{\tiny EH}} $ being the Einstein-Hilbert AdS action and $\mathcal{L}_{\mbox{\tiny NG}}$ being a Nambu-Goto (NG) brane action, that is, $\mathcal{L}_{\mbox{\tiny NG}}=T_p\sqrt{\gamma_{\mbox{\tiny NG}}}$. The parameter that controls the backreaction is just $t=T_pL^{p+1}$. For $t<<1$ we can make a perturbative expansion of the metric as
\begin{equation}
    g_{\mu \nu}=g^{(0)}_{\mu \nu}+g^{(1)}_{\mu \nu}+\mathcal{O}(t^2)~,
\end{equation}
where $g_{\mu \nu}$ is the backreacted metric and $g^{(0)}_{\mu \nu}$ satisfies the Einstein equations. Now we can expand \eqref{bulk plus brane action} as follows:
\begin{equation}
    I[g]=I_{\mbox{\tiny EH}}[g^{0}]+I_{\mbox{\tiny NG}}[g^{0}]+ g^{(1)}_{\mu\nu}\cancel{\frac{\delta I_{\mbox{\tiny EH}}[g^{0}] }{\delta g_{\mu\nu}}}+\mathcal{O}(t^2)~.
\end{equation}
At zero order, $\mathcal{O}(t^0)$, the only term is $I_{\mbox{\tiny EH}}[g^{0}]$. At first order, $\mathcal{O}(t)$, we have  two contributions. A crucial step is to note that the second of these contributions vanishes because of the equations of motion and therefore, to first order, the EE can be computed simply by using \eqref{LM EE}, where  the action is \eqref{bulk plus brane action} evaluated in the non-backreacted metric. Doing so we get

\begin{equation}
    S=S^{(0)}+S^{(1)},
\end{equation}
where the superindices take account of the order in the $t$ expansion. Of course, $S^{(0)}$ will be just the \eqref{RT formula},  and $S^{(1)}$ is
\begin{equation}
    S^{(1)}= \partial_n I_{\mbox{\tiny NG}}|_{n=1}=\left[\int_{0}^{r_{\epsilon}} dr d^p y \partial_n \mathcal{L}_{\mbox{\tiny NG}}+ \partial_n I_{ct}\right]_{n=1}
\end{equation}
Here, $r_{\epsilon}$ is the radial cut-off. 

We can interpret the derivative with respect to $n$, evaluated at $n=1$, as a first-order variation, denoted by $\delta_n$. As shown in \cite{Lewkowycz:2013nqa,Karch:2014ufa}, the derivative of the brane Lagrangian can be rewritten via integration by parts, rendering the integrand proportional to the equations of motion:

\begin{equation}
  \delta_n \mathcal{L}_{\mbox{\tiny NG}}=\frac{\delta\mathcal{L}_{\mbox{\tiny NG}}}{\delta g_{\mu \nu}}\delta_n g_{\mu \nu}+ \cancel{\frac{\delta\mathcal{L}_{\mbox{\tiny NG}}}{\delta X_{\mu }}}\delta_n X_{\mu }+\partial_{\mu}\Theta^{\mu}[\delta X]. \label{derivative n of brane lagrangian}
\end{equation}
Here, $X^{\mu}$ are the embedding functions of the brane and $\Theta^{\mu}[\delta X]$ is a boundary term. Notice that there are no boundary terms concerning the bulk metric because the NG Lagragian does not involve  derivatives of it. The second term has been canceled because the embedding functions of the brane are on shell for $n=1$.

It is worth mentioning that for the Einstein AdS action the same procedure could be applied and the whole contribution for the EE will come from the boundary term located at $r=0$, as seen in \eqref{Action in replica orbifold}. However, because the brane action   just involves the volume form, there is no additional contribution from $r=0$. Therefore the EE up to first order in $t$ is just
\begin{equation}
    S=\frac{A_{\text{min}}}{4G}+\int dr d^p y  \frac{\delta \mathcal{L}_{\mbox{\tiny NG}}}{\delta g_{\mu\nu}}\delta_n g_{\mu\nu}+\int_{r=r_{\epsilon}} d^p y\frac{\delta\mathcal{L}^{ct}_{\mbox{\tiny NG}}}
    {\delta g_{\mu\nu}}\delta_n g_{\mu\nu}, \label{brane entropy}
\end{equation}
where the same procedure of Eq.\eqref{derivative n of brane lagrangian} has been applied to the  brane counterterm Lagrangian, obtaining the variation of it with respect to changes in the bulk metric and evaluating at the cutoff $r_{\epsilon}$. There is also an additional contribution from the counterterm arising from the variation of the Lagrangian with respect to the embedding functions $X^\mu$. However, it is canceled by the boundary term of \eqref{derivative n of brane lagrangian}, because the brane counterterms are defined in such a way that the action is stationary.

The key takeaway from this general discussion is that we can determine the contribution of the probe brane to the EE simply by analyzing the variation of the NG action concerning changes in the bulk metric with respect to 
$n$. There is no need to compute the backreaction, and all that is required is the appropriate brane embedding for the $n=1$ case.

\subsection{EE for the string dual to the shifted quark-antiquark configuration}\label{section 4.3}

Our starting point for these computations will be \eqref{brane entropy}, particularized for the string action:

\begin{equation}
S = S^{(0)}+S^{(1)}=\frac{A_{\text{min}}}{4G} + \int dr \int d\tau \frac{\delta \mathcal{L}_{s}}{\delta g_{\mu\nu}}\delta_n g_{\mu\nu} + \int_{r=r_{\epsilon}} d\tau \frac{\delta\mathcal{L}_s^{ct}}{\delta g_{\mu\nu}}\delta_n g_{\mu\nu}.\label{RT plus string EE}
\end{equation}
Here, $\mathcal{L}_{s} = T\sqrt{\gamma}$, where $T$ is the tension of the string and $\gamma_{ab} = \partial_a X^{\mu} \partial_b X_{\mu}$ is the induced worldsheet metric. Additionally, $\mathcal{L}_s^{ct} = TL\sqrt{\gamma_\epsilon}$ denotes the counterterm Lagrangian for the string action.

To perform the explicit computation, we will work on the $n$-fold cover of the  bulk space. For a spherical entangling surface, there is a natural coordinate system describing this space \cite{Karch:2014ufa}, which is a generalization of the AdS hyperbolic coordinates given in \eqref{hyperbolic AdS metric}:
\begin{equation}
ds^2_n = L^2 \left[f_n(u) d\tau^2 + \frac{du^2}{f_n(u)} + u^2 (d\rho^2 + \sinh^2(\rho) (d\theta^2 + \sin^2(\theta) d\phi^2))\right] , \label{Euclidean hyperbolic AdS black holes}
\end{equation}
where $f_n(u)=u^2-1-c_h u^{-2}$ and $c_h\equiv u_h^4-u_h^2$. Equation \eqref{Euclidean hyperbolic AdS black holes} is also the Euclidean version of a family of topological AdS black holes discussed in \cite{Emparan:1999gf}. In terms of the replica parameter $n$, the location of the horizon is at $\displaystyle{u_h=\frac{1}{4n}[1+\sqrt{8n^2+1}]}$.  It is worth noticing that for the $n=1$ case, we have $u_h=1$ and therefore  $c_h=0$, recovering the metric of pure AdS in hyperbolic coordinates \eqref{hyperbolic AdS metric}.

Because we are going to make use of the embedding function  \eqref{string dp r} to compute the entanglement entropy, it will be better to perform the transformations of \eqref{Hyperbolic rho}-\eqref{Hyperbolic theta}. Then, we will be working on the $n$-fold generalization of the AdS double polar coordinates:
\begin{equation}
\bar{g}^{(n)}_{\mu\nu}d\bar{x}^{\mu}d\bar{x}^{\nu} = L^2 \left[f_n(u) d\tau^2 + \frac{du^2}{f_n(u)} + \frac{u^2}{r^2} (dr^2 + dy^2+y^2d\phi^2)\right]~.
\label{Euclidean hyperbolic AdS black holes DP}
\end{equation}
In these new coordinates, the contribution of the string to the EE is
\small
\begin{equation}
    S^{(1)}=-n^2\partial_nu_h|_{n=1}T\int_{u=u_h} d\tau  \sqrt{\bar{\gamma}}+ T\int_{u_h}^{u_\infty} du \int_{0}^{2\pi} d\tau \frac{\delta\sqrt{\bar{\gamma}}}{\delta \bar{g}_{\mu\nu}}\delta_n \bar{g}_{\mu\nu} + TL\int_{u=u_{\infty}} d\tau \frac{\delta\sqrt{\bar{\gamma}_{\epsilon}}}{\delta \bar{g}_{\mu\nu}}\delta_n \bar{g}_{\mu\nu},\label{EE string}
\end{equation}
\normalsize
where the first term appears because the lower limit of the integral depends on $n$. As you can notice, this term is evaluated at the fixed value $u=u_h$, which corresponds to the location of the RT surface. Therefore, it is the contribution to the EE coming from the intersection between this surface and the string worldsheet.

Now we will  obtain all the ingredients that we need to compute the entanglement entropy. Let us start with the variation with respect to $n$ for this metric. This is straightforward, since the only components of the metric depending on $n$ are $g_{00}$ and $g_{11}$:
\begin{equation}
\delta_{n}\bar{g}_{\mu\nu}=\begin{pmatrix}
\frac{2L^2}{3u^2} & 0 & 0 & 0 & 0\\
0 & -\frac{2L^2}{3u^2(u^2-1)^2}& 0 & 0 & 0\\
0 & 0 & 0 & 0 & 0\\
0& 0 & 0 & 0 & 0\\
0 & 0 & 0 & 0 & 0
\end{pmatrix},
\end{equation}

To obtain the  embedding function $\bar{X}^{\mu}(\tau,u)$ of the string in these coordinates, we will start with the embedding given in \cite{Xiao:2008nr}, and then perform the coordinate transformation. We have that
\begin{equation}
    X^{\mu}(x^0,z)=(x^0,z,x^1(x^0,z),0,0)
\end{equation}
where $x^1(x^0,z)$ is defined by \eqref{string in AdS}. We know that $\partial_a \bar{X}^{\mu}$ transforms as a vector under diffeomorphisms,
\begin{equation}
   \partial_a \bar{X}^{\mu}=\frac{\partial \bar{x}^{\mu}}{\partial x^{\nu}} \partial_aX^{\nu},
\end{equation}
where $\bar{x}^{\mu}$ denotes the Poincaré coordinates and $x^{\mu}$ the AdS double polar coordinates. We obtain 
\begin{align}
     \partial_0 \bar{X}^{\mu}&=(1,0,\Dot{r}(\tau,u),0,0),\label{partial 0 embedding}\\
     \partial_1 \bar{X}^{\mu}&=(0,1,r'(\tau,u),0,0).\label{partial 1 embedding}
\end{align}
Here, $r(\tau,u)$ is the embedding function of \eqref{string dp r}, and $\Dot{r}(\tau,u)$ and $r'(\tau,u)$ are its partial derivatives with respect to $\tau$ and $u$ respectively.  Because the worldsheet metric is just $\bar{\gamma}_{ab}=\partial_a\bar{X}^{\mu}\partial_b\bar{X}_{\mu}$, we have 

\begin{equation}
\bar{\gamma}_{ab}=\begin{pmatrix}
 L^2(u^2-1)+\frac{L^2u^2(\Dot{r}(\tau,u))^2}{r^2(\tau,u)}& -\frac{L^2u^2r'(\tau,u)\Dot{r}(\tau,u)}{r^2(\tau,u)}\\
-\frac{L^2u^2r'(\tau,u)\Dot{r}(\tau,u)}{r^2(\tau,u)} & \frac{L^2}{u^2-1}+\frac{L^2u^2(r'(\tau,u))^2}{r^2(\tau,u)} \\
\end{pmatrix},
\end{equation}
and therefore the determinant is
\begin{equation}
  \bar{\gamma}=L^4\frac{r^2(\tau,u)(u^2-1)+u^2(r'(\tau,u))^2(u^2-1)^2+(\Dot{r}(\tau,u))^2u^2}{r^2(\tau,u)(u^2-1)}\:.\label{determinant r}
\end{equation}

Notice that when $h=0$ (and therefore $r(\tau,u)=b$), we obtain that $\bar{\gamma}=L^4$ does not depend on the worldsheet coordinates and, therefore, the calculation of the EE would be greatly simplified. On the other hand, if  $h$ is different from zero, then after replacing  \eqref{string dp r} into \eqref{determinant r}  we get
\begin{equation}
    \bar{\gamma}=L^4\frac{2b^2u^2}{2b^2u^2-h^2(u^2+1)+h^2(u^2-1)\cos(2\tau)}~.
\end{equation}

With all these ingredients, now it is possible to compute  the first term of \eqref{EE string}:
\small
\begin{align}
   S^{(1)}_1= - n^2\partial_nu_h|_{n=1}T\int_{u=1} d\tau  \sqrt{\bar{\gamma}} &=\frac{T}{3}\int_{0}^{2\pi} d\tau\left(\sqrt{L^4\frac{2b^2u^2}{2b^2u^2-h^2(u^2+1)+h^2(u^2-1)\cos(2\tau)}}\right)_{u=1}\nonumber\\
    &=\frac{TL^2}{3}\int_{0}^{2\pi} d\tau\left(\sqrt{\frac{b^2}{b^2-h^2}}\right)=\frac{2\pi T L^2}{3\sqrt{1-h^2/b^2}}=\frac{\sqrt{\lambda}}{3\sqrt{1-h^2/b^2}}~,
    \label{EE string 1}
\end{align}
\normalsize
where the result  $- n^2\partial_nu_h|_{n=1}=\frac{1}{3}$ has been used. The holographic correspondence tells us that the string tension is related to the 't Hooft coupling $\lambda$ by the equation $T=\frac{\sqrt{\lambda}}{2\pi L^2}$ \cite{Maldacena:1997re,Maldacena:1998im}. This relation has been used in the last equality of the previous equation. Notice that for $h=0$ we get that the contribution of this term to the EE is just $\frac{\sqrt{\lambda}}{3}$. This is, in fact,  what has been obtained previously in the literature for the non-shifted case \cite{Jensen:2013lxa}, so we conclude that $S^{(1)}_2=0$ in that case.

For the second term on the right-hand side of  \eqref{EE string}, we should take into account that the integrand can be expressed as
\footnotesize
\begin{align}
    &I=\frac{\delta\sqrt{\bar{\gamma}}}{\delta \bar{g}_{\mu\nu}}\delta_n \bar{g}_{\mu\nu}=\frac{1}{2}\sqrt{\bar{\gamma}}\bar{\gamma}^{ab}\partial_a \bar{X}^{\mu}\partial_b\bar{X}^{\nu}\delta_n \bar{g}_{\mu \nu} \nonumber\\
    &=\sqrt{\frac{(u^2-1)}{(r^2(\tau,u)(u^2-1)+u^2(r'(\tau,u))^2(u^2-1)^2+(\dot{r}(\tau,u))^2u^2)}}\frac{L^2((r'(\tau,u))^2(u^2-1)-(\dot{r}(\tau,u))^2)}{3(u^2-1)^2r(\tau,u)}\nonumber\\
   &=\frac{h^2L^2}{3b}\sqrt{\frac{1}{b^2u^2-h^2(\cos^2(\tau)+u^2\sin^2(\tau))}}\frac{\cos^2(\tau)-u^2\sin^2(\tau)}{u^3(u^2-1)}~. \label{Secondintegrand}
\end{align}
\normalsize 
In the last line of this procedure, we have used the explicit embedding  given in \eqref{string dp r}. As expected, for the case when $h=0$, this term goes to zero and therefore does not contribute to the entanglement entropy. For the more general case when $h \neq 0$, we can carry out a Taylor expansion of the integrand around  $\frac{h}{b} < 1$ :

\footnotesize
\begin{align}
 &I=\frac{TL^2}{3}\left(\frac{h}{b}\right)^2\sqrt{\frac{1}{u^2-\left(\frac{h}{b}\right)^2(\cos^2(\tau)+u^2\sin^2(\tau))}}\frac{\cos^2(\tau)-u^2\sin^2(\tau)}{u^3(u^2-1)}\nonumber\\
&= \frac{TL^2}{3}\sum_{n=0}^{\infty}\prod^{n}_{i=1}(2i-1)\frac{1}{2^n n!}\left(\frac{h}{b}\right)^{2(n+1)}\left[\frac{\left(\cos^2(\tau)+u^2\sin^2(\tau)\right)^n}{u^{2n}}\right] \frac{\cos^2(\tau)-u^2\sin^2(\tau)}{u^4(u^2-1)} 
 \nonumber\\
 &= \frac{TL^2}{3}\sum_{n=0}^{\infty}\prod^{n}_{i=1}(2i-1)\frac{1}{2^n n!}\left(\frac{h}{b}\right)^{2(n+1)}\left[\frac{\left((\frac{1-u^2}{u^2})\cos^2(\tau)+1\right)^n\left((\frac{1+u^2}{u^2})\cos^2(\tau)-1\right)}{u^{2}(u^2-1)}\right] 
 \nonumber\\
 &= \frac{TL^2}{3}\sum_{n=0}^{\infty}\sum_{k=0}^{n}\prod^{n}_{i=1}(2i-1)\binom{n}{k}\frac{1}{2^n n!}\left(\frac{h}{b}\right)^{2(n+1)}\left[\frac{(1-u^2)^{k-1}}{u^{2(k+1)}}\cos^{2k}(\tau)-\frac{(1+u^2)(1-u^2)^{k-1}}{u^{2(k+2)}}\cos^{2(k+1)}(\tau)\right]~. \label{Integrand 1}
\end{align}
\normalsize
 In the last equality, we have applied a binomial expansion to $\left((\frac{1-u^2}{u^2})\cos^2(\tau)+1\right)^n$. Taking into account the  result $\int_0^{2\pi}d\tau\cos(\tau)^{2k}=\frac{2\pi}{2^{2k}}\binom{2k}{k}=\frac{2\pi}{2^{2k}}\frac{(2k)!)}{(k!)^2}$, we can integrate \eqref{Integrand 1} with respect to $\tau$:
\footnotesize
 \begin{equation}
     \int_{0}^{2\pi}d\tau I=\frac{TL^2}{3}\sum_{n=0}^{\infty}\sum_{k=0}^{n}\prod^{n}_{i=1}(2i-1)\binom{n}{k}\frac{1}{2^n n!}\left(\frac{h}{b}\right)^{2(n+1)}\frac{2\pi}{2^{2k}}\binom{2k}{k}\frac{(1-u^2)^{k-1}}{u^{2(k+1)}}\left[1-\frac{(1+u^2)}{2^2u^2}\frac{(2k+1)(2k+2)}{(k+1)^2}\right]~.
 \end{equation}

\normalsize 
Now, to perform the integration with respect to $u$, we will make a binomial expansion of $(1-u^2)^{k-1}$. We should be careful for the case when $k=0$, due to the fact that in that case the exponent will be negative. Therefore, we separate it from the sum over $k$:

\begin{align}
     & \int_{0}^{2\pi}d\tau I=\frac{TL^2}{3}\sum_{n=0}^{\infty}\prod^{n}_{i=1}(2i-1)\frac{1}{2^n n!}\left(\frac{h}{b}\right)^{2(n+1)}2\pi\left[-\frac{1}{2u^4}\right.\\
      &\left.+\sum_{k=1}^{n}\sum_{j=0}^{k-1}\binom{n}{k}\binom{2k}{k}\binom{k-1}{j}\frac{(-1)^j}{2^{2k}}u^{2(j-(k
     +1))}\left[\frac{1}{2(k+1)}-\frac{(2k+1)(2k+2)}{2^2u^2(k+1)^2}\right]\right]~.\nonumber
\end{align}
    
Finally, we integrate over $u$, and after taking the limits from $u_h=1$ to $u_\infty\rightarrow\infty$, we obtain
\begin{align}
    S^{(1)}_2= & \int_{1}^{\infty }du\int_{0}^{2\pi}d\tau I=-\frac{TL^2}{3}\sum_{n=0}^{\infty}\prod^{n}_{i=1}(2i-1)\frac{1}{2^n n!}\left(\frac{h}{b}\right)^{2(n+1)}2\pi\left[\frac{1}{6}\right.\label{EE string 2}\\
      &\left.+\sum_{k=1}^{n}\sum_{j=0}^{k-1}\binom{n}{k}\binom{2k}{k}\binom{k-1}{j}\frac{(-1)^j}{2^{2k}}\left[\frac{1-k(1-2j+2k)}{(3-2j+2k)(-1+2j-2k)(k+1)}\right]\right]~.\nonumber 
\end{align}

\begin{figure}[h!]
    \centering
    \begin{subfigure}{0.45\textwidth}
        \centering
        \includegraphics[width=\textwidth]{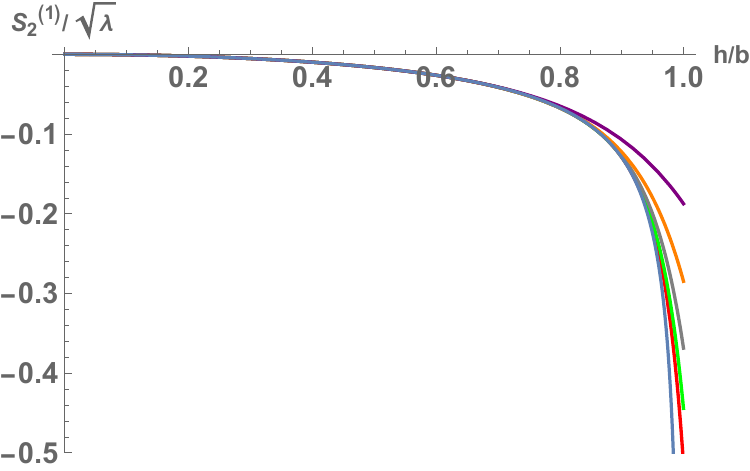}
        \caption{}
    \end{subfigure}
    \hfill
    \begin{subfigure}{0.45\textwidth}
        \centering
        \includegraphics[width=\textwidth]{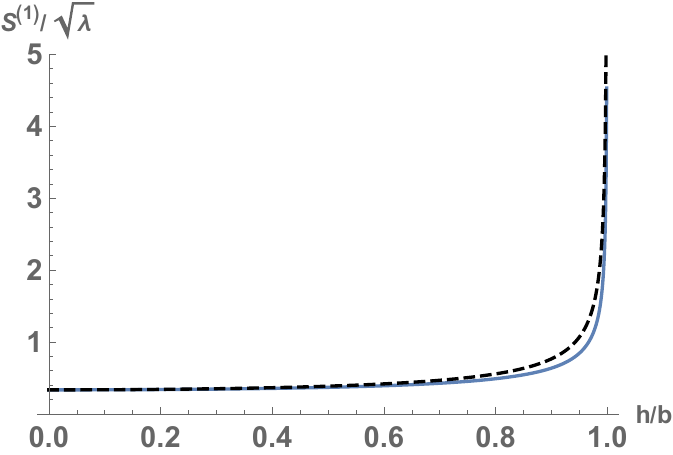}
        \caption{}
    \end{subfigure}
    \caption{(a) Here we have the plot of \eqref{EE string 2} as a function of $h/b$. Each of the curves corresponds to a different value of truncation of the sum, ranging from $n = 5$ to $n = 25$. The blue line represents the result obtained by numerically integrating \eqref{Secondintegrand}. (b) 
The dashed black line is the plot of \eqref{EE string 1} and the blue line is the numerical result obtained for the complete contribution of the string to the  EE \eqref{EE string}. The discrepancy between the two is due to the contribution that (a) has on the final result.}
    \label{fig:4}
\end{figure}

In Fig.~\ref{fig:4}a, we can see this result depicted, where we have plotted the graphs as a function of $h/b$ for different orders of truncation in the sum over $n$. It can be observed that as we increase the order of truncation in $n$, we correctly approach the result obtained numerically (blue curve).

Now it is time to  look at the last term of Eq.~(\ref{EE string}), which involves the countertem Lagrangian. The metric $\bar{\gamma}_{\epsilon}$  is obtained from $\bar{\gamma}_{ab}$ just by taking the limit $u\rightarrow \infty$. Of course, this new metric has just one component,

\begin{equation}
    \bar{\gamma}_{\epsilon}=\lim_{u\rightarrow\infty}\frac{L^2(u^2-1)\left(2b^2u^2-h^2(1+\cos(2\tau))\right)}{2\left((b^2-h^2)u^2+h^2(u^2-1)\cos^2(\tau)\right)}~.
\end{equation}

Now, the integrand is
\begin{align}
   & \frac{\delta\sqrt{\bar{\gamma}_{\epsilon}}}{\delta \bar{g}_{\mu\nu}}\delta_n \bar{g}_{\mu\nu}=\frac{1}{2} \sqrt{\bar{\gamma}_{\epsilon}}\bar{\gamma}_{\epsilon}^{-1}\delta_0\bar{X}^{\mu}_{\epsilon}\delta_0 \bar{X}^{\nu}_{\epsilon}\delta_n g_{\mu\nu}\nonumber\\
    &= \lim\limits_{u\rightarrow\infty}\frac{L}{3}\frac{1}{u^2}\sqrt{\frac{\left((b^2-h^2)u^2+h^2(u^2-1)\cos^2(\tau)\right)}{(u^2-1)\left(b^2u^2-h^2\cos(\tau)^2)\right)}}=0~,\label{EE string 3}
\end{align}
where we have used the result $\delta_0\bar{X}^{\mu}_{\epsilon}\delta_0 \bar{X}^{\nu}_{\epsilon}\delta_n \bar{g}_{\mu\nu}=\lim\limits_{u\rightarrow\infty}\frac{2}{3}\frac{L^2}{u^2}$. The vanishing result in \eqref{EE string 3} is expected, since the other two terms contributing to the EE are finite.

Using \eqref{EE string 1}-\eqref{EE string 3}, we see that the contribution from the accelerating string to the EE is
\begin{align}
     S^{(1)}&=S^{(1)}_1+S^{(1)}_2=\frac{\sqrt{\lambda}}{3}\left[\frac{1}{\sqrt{1-\left(\frac{h}{b}\right)^2}}-\sum_{n=0}^{\infty}\prod^{n}_{i=1}(2i-1)\frac{1}{2^n n!}\left(\frac{h}{b}\right)^{2(n+1)}\left[\frac{1}{6}\right.\right.\label{FinalresultEE}\\
      &\left.\left.+\sum_{k=1}^{n}\sum_{j=0}^{k-1}\binom{n}{k}\binom{2k}{k}\binom{k-1}{j}\frac{(-1)^j}{2^{2k}}\left(\frac{1-k(1-2j+2k)}{(3-2j+2k)(-1+2j-2k)(k+1)}\right)\right]\right]~.\nonumber 
\end{align}
Here, it is clear that when  $h = 0$, we recover the $\sqrt{\lambda}/3$ result of the non-shifted case \cite{Jensen:2013ora}. The behavior of \eqref{FinalresultEE} for $\frac{h}{b} \ll 1$ is
\begin{equation}
    S^{(1)}=\frac{\sqrt{\lambda}}{3}\left(1+\frac{1}{3}\left(\frac{h}{b}\right)^2+ \mathcal{O}\left(\left(\frac{h}{b}\right)^4\right)\right).
\end{equation}

On the other hand, the EE diverges when $\frac{h}{b}\to 1$, as anticipated in Section~\ref{section 3.3}. Near this limit, the EE behaves as
\begin{equation}
S^{(1)} = \frac{\sqrt{\lambda}}{3} \left( \frac{1}{\sqrt{2(1 - h/b)}} + \mathcal{O}\left(\sqrt{1 - h/b}\right) \right),
  \label{EE divergence}
\end{equation}

Notice that $S_1^{(1)}$ is the dominant contribution to the defect entanglement entropy $S^{(1)}$ for all values of the dimensionless parameter $h/b$, and that the positive divergent behavior as $h/b \to 1$ originates entirely from this term. In the worldsheet interpretation, $S_1^{(1)}$ would naturally correspond to an “area” term associated with the worldsheet entangling surface. As seen in (\ref{EE string 1}), this term is simply the Nambu–Goto Lagrangian density evaluated at $u = u_h$, integrated over $\tau$. The second term, $S_2^{(1)}$, is subdominant and arises from the convolution between $\frac{\delta\sqrt{\bar{\gamma}}}{\delta \bar{g}_{\mu\nu}}$, which is proportional to the spacetime stress tensor $T^{\mu\nu}$ generated by the string, and the metric variation $\delta_n \bar{g}_{\mu\nu}$, which contributes only to the $\tau\tau$ and $uu$ components. Therefore, $S_2^{(1)}$ can be interpreted as encoding  energy density ($T^{\tau\tau}$) and momentum flux ($T^{uu}$).

Although we have performed this computation on an AdS$_5$ background, its generalization to arbitrary dimension is straightforward. For AdS$_{d+1}$, we obtain

\begin{align}
  S(d)^{(1)}&=\frac{\sqrt{\lambda}}{d-1}\left[\frac{1}{\sqrt{1-\left(\frac{h}{b}\right)^2}}-\sum_{n=0}^{\infty}\prod^{n}_{i=1}(2i-1)\frac{1}{2^n n!}\left(\frac{h}{b}\right)^{2(n+1)}\left[\frac{1}{2(d-1)}\right.\right.\label{FinalresultEEd}\\
      &\left.\left.+\sum_{k=1}^{n}\sum_{j=0}^{k-1}\binom{n}{k}\binom{2k}{k}\binom{k-1}{j}\frac{(-1)^j}{2^{2k}}\left(\frac{1-k(d-3-2j+2k)}{(d-1-2j+2k)(-d+3+2j-2k)(k+1)}\right)\right]\right]~.\nonumber 
\end{align}
Once again, in the limit $h/b \to 0$, it reduces to previously known results (see Appendix B of \cite{Lewkowycz:2013laa}). Notice that the nature of the divergence is independent of the background dimension, scaling universally as $\epsilon^{-1/2}$, where $\epsilon = h - b$ is the distance between the quark and the entangling surface. This is consistent with the fact that the EE characterizes information associated with the defect, and therefore its divergence should be governed by the dimensionality of the defect itself.

As mentioned in Section~\ref{introduction}, the case that makes contact with a double holographic setup is $d=2$. In light of our results, it is tempting to draw an analogy between the two components of the defect entropy and the corresponding terms that appear in the definition of generalized gravitational entropy in JT gravity~\cite{Almheiri:2019hni,Almheiri:2019psf}. In this view, $S_1^{(1)}$ would play the role of the worldsheet RT contribution, analogous to the dilaton term ~\cite{Jafferis:2019wkd}, while the subleading term $S_2^{(1)}$ could be associated with the EE of fields living on the worldsheet gravitational theory. It is important to emphasize, however, that as plausible as this interpretation might seem, thus far there is no concrete reason for it to hold, since from the bulk point of view we are just computing the change in the area of the RT surface due to the backreaction of the string. Certainly we do not claim that the worldsheet dynamics is governed by JT gravity, nor that the present setup realizes island physics in a literal sense.\footnote{See footnote~\ref{islandfoot}.} Rather, the analogy offers a useful organizing principle and suggests directions for future work aimed at understanding whether a controlled, intrinsic worldsheet description can be made precise and applied to computations involving Hawking radiation.

\acknowledgments

We thank Ignacio Araya, Daniel Ávila, Mariano Chernicoff  and Sergio Patiño for useful conversations, and Juan Pedraza for valuable discussions. We are also grateful to Luis León for comments on the manuscript. Our work was partially supported by DGAPA UNAM grant IN116823. 

\appendix

\section{A note on the EE for  \texorpdfstring{$h>b$}{h>b}} \label{Appendix A}

For the case where $h > b$,  after switching to Euclidean signature and performing the double polar transformations, we must consider that the circular Wilson loop is composed of two branches \footnote{For the case $h \leq b$, we only considered the positive branch, as the other one remains negative for all values of $\tau$.},
\begin{equation}
\text{r}_{\pm}(\tau) = h \cos(\tau) \pm \sqrt{b^2 - h^2 \sin(\tau)^2}.
\end{equation}
Here, the domain of the $\tau$ coordinate is constrained to the range
\begin{equation}
    -\frac{b}{h} \leq \sin(\tau) \leq \frac{b}{h} \label{constraint CFT tau}
\end{equation}
 to ensure that $\text{r}(\tau)$ remains real. We account for both branches because $\text{r}(\tau)$ is double-valued, capturing all points along the circumference.

As we already know, the dual description of the infinitely heavy accelerating  $q$-$\bar{q}$ pair is an accelerating string attached to the boundary of AdS described by \eqref{string in AdS}. After performing the Wick rotation and  the double polar AdS transformations, the string worldsheet will be described by
\begin{equation}
r_{\pm}(\tau,u) = h\sqrt{1-1/u^2} \cos(\tau) \pm \sqrt{b^2 - h^2 +h^2(1-1/u^2)\cos(\tau)^2}\:\:.
\end{equation}
It is important to note that for the solution to be real we should restrict the values of $\tau$ at each value of $u$,
\begin{equation}
  -\sqrt{\left(\frac{b}{h}\right)^2-\frac{1}{u^2-1}} \leq \sin(\tau) \leq \sqrt{\left(\frac{b}{h}\right)^2-\frac{1}{u^2-1}}.\label{constraint string tau}
\end{equation}
We see here that when  $u\rightarrow \infty$  the constraint \eqref{constraint CFT tau} is recovered.

At first glance, the computation of the EE should be similar to the $h\leq b$ case, with the only difference being that the constraint from \eqref{constraint string tau} should be implemented in the limits of integration of \eqref{EE string}. However, there is an additional non-trivial task to perform if we wish to fully obtain the EE.

As mentioned earlier, the first term in \eqref{EE string} appears because the limit of integration for $u$ depends on the replica parameter $n$, and we already know that this dependence takes the form $ u_h = \frac{1}{4n} \left(1 + \sqrt{8n^2 + 1}\right)$. This  $n$-dependence of the integration limit remains unchanged as long as $h\leq b$. However, for $h>b$, the generalization of \eqref{constraint string tau} containing the $n$-dependence would require first obtaining the string embedding in the $n$-fold cover of AdS, which is a difficult task that has not yet been addressed in the literature.

\newpage
\bibliography{Bibliography.bib}
\bibliographystyle{./utphys}

\end{document}